\documentstyle[apjpt4]{article} 

\newcommand{\etal}{} \def\etal/{et al.} 
\newcommand{\kms}{} \def\kms/{km s$^{-1}$} 
\newcommand{\degre}{} \def\degre/{$^\circ$}
\newcommand{\halpha}{} \def\halpha/{H$\alpha$} 
\newcommand{\hbeta}{} \def\hbeta/{H$\beta$} 
\newcommand{\flux}{} \def\flux/{erg cm$^{-2}$ s$^{-1}$} 
\newcommand{\allo}[3]{\ion{#1}{#2}$\lambda$\-#3} 
\newcommand{\forb}[3]{[\ion{#1}{#2}]$\lambda$\-#3} 
 
\newcommand{\dforb}[4]{[\ion{#1}{#2}]$\lambda\lambda$\-#3,#4} 

\newcommand{\mvelo}[1]{$<V>_{#1}$} 
\newcommand{\diab}{} \def\diab/{Diabolo} 
 
\begin{document} 
\twocolumn 
 
\title{Morphology and kinematics of Planetary Nebulae\\ 
II. A \diab/ model for NGC~3132 
} 

\author{H. Monteiro\altaffilmark{1}, C. Morisset\altaffilmark{1,2},
R. Gruenwald\altaffilmark{1}, \and S. M. Viegas\altaffilmark{1} } 

\altaffiltext{1}{Instituto Astron\^omico e Geof\'{\i}sico, USP,
Avenida Miguel Stefano, 4200 CEP 04301-904 S\~ao Paulo, SP, Brazil}
\altaffiltext{2}{Laboratoire d'Astronomie Spatiale, Traverse du Siphon, Les 
Trois Lucs, 13012 Marseille, France }

\begin{abstract} 

We use a 3D photoionization modeling tool to study the
morpho-kinematic properties of the Planetary Nebula (PN) NGC~3132. We
show that it is 
possible to reproduce the low resolution observations (spectra and 
images) with an ellipsoidal shell. 
However, high resolution observations, as those showing a density variation 
along the nebula and the
\forb{O}{3}{5007} velocity profiles, definitively rule out this
description.
We show that a bipolar Diabolo shape with a 40\degre/ rotation
of the symmetry axis relative to the line of sight successfully
reproduces the observed images, as well as the high resolution observations. 

\end{abstract} 
 
\keywords{Methods: numerical -- planetary nebulae: NGC~3132} 
 
\section{Introduction} 
 
The determination of the three-dimensional matter distribution in planetary 
nebulae is essential for a knowledge of the nebula ejection mechanism.
A precise understanding of the nebula geometry which produces the observed 
morphology can also provide constraints on stellar evolution theories.
The geometry and the density 
distribution of PNe are generally derived from line imaging, assuming that 
a surface brightness enhancement corresponds to a density enhancement. 
After a first descriptive study of the morphology of PNe (Curtis 1918), 
more detailed studies indicate two different interpretations for the 
observed morphology of these objects:
a) the different observed morphologies are due to different projections
of a same common basic structure (Minkowski \& Osterbrock 1960, Khromov \& 
Kohoutek 1968); b) the different types in which PNe can be classified 
(bipolar, elliptical, etc.) correspond to different evolutionary stages 
(Balick 1987). In this second approach, the morphology can also be related 
to the 
nebular and stellar properties, particularly the mass of the progenitor star
(e.g. Peimbert \& Torres-Peimbert 1983, Calvet \& Peimbert 1983, 
Stanghellini, Corradi \& Schwarz 1993, Corradi \& Schwarz 1995). This 
point of view is corroborated by the studies of Zhang \& Kwok (1998). These
authors, assuming a three-dimensional ellipsoidal shell with density variations
and a complete absorption of the ionizing photons,
which are reconverted in H recombination photons, 
produce simulated radio and H$\alpha$ images which are compared to the 
observations.
The visual appearence of PNe were also compared with models obtained for 
ionization-bounded prolate shells by Masson (1990). Assuming 
an ellipsoidal shell with constant thickness and that the 
ionizing radiation is geometrically diluted, Masson (1990) compared
theoretical and observed H$\alpha$ images.
However, a given emission line image results from 
the sum of line intensities along the line of sight. 
The exact location of the regions which produce each line depends 
on the gas density distribution, but also on the 
ionization distribution, which, in turn, depends on the characteristics 
of the ionizing star and on the nebular gas abundance. Consequently, 
different emission lines provides information on different 
regions of the nebula, and only a consistent photoionization model 
applied to a realistic gas structure can give an accurate density distribution 
on the nebula.

In this paper we show that the geometry of a PN derived by fitting  
the observed image of a line emission and/or a line intensity 
ratio is misleading, since different geometries can reproduce the 
same observational data. Other constraints, coupled with a detailed 3D 
photoionization model, must be used for better defining the geometry, as 
illustrated by the planetary nebula NGC~3132.

Based on emission line imaging, an ellipsoidal geometry for NGC~3132 
is assumed by various authors. Such a geometry is then suggested in a
first approach. Studies similar to that of Masson (1990) using a prolate 
shell could explain the observed morphology of NGC~3132. Notice that 
the ellipsoidal shell model of 
Zhang \& Kwok (1998) also reproduces the general trends of the observed 
H$\alpha$ image of NGC~3132. 
Using a geometry defined by two concentric ellipsoids and a
3D photoionization code,
B\"assgen, Diesch \& Grewing (1990, hereafter \cite{B90})  
fit the observed [\ion{O}{3}] and [\ion{N}{2}] line images, 
as well as the emission line spectra taken through four slits. 

Spatio-kinematic models for NGC~3132 
were proposed by Sahu \& Desai (1986, hereafter \cite{SD86}) to explain
the observed asymmetric double-peaked \forb{O}{3}{5007} line profiles in 
five positions. 
\cite{SD86} assert that an ellipsoidal model 
with velocity and density asymmetries reproduces the observed expansion 
velocities and correctly explains the nature of the profile asymmetries.
However, in their paper, only the fit to the line profile at the central 
position is shown.  
Line profiles, as a powerful tool for studying the nebula geometry, 
are discussed by Morisset, Gruenwald \& Viegas (2000, hereafter MGV2000).

As it will be shown below, all the models assuming an ellipsoidal emitting 
shell lead to the same problems: neither 
can reproduce the low central density observed by \cite{SD86} and 
\cite{J88} nor the observed asymmetric double-peaked profiles 
in regions far from the center of the nebula. 
After a description of the available
observations of NGC~3132 (\S \ref{sec-obs}), we show that  
ellipsoidal shell models do not reproduce all the available
observations (\S \ref{sec-bass}). A ``\diab/'' nebula  
successfully explaining  all the main observational trends is presented in 
\S \ref{sec-diab}. The conclusions are outlined in \S \ref{sec-concl}. 

\section{Observational data for NGC~3132} 
\label{sec-obs} 

A consistent photoionization model giving a realistic description  
of a PN has to account for all the observational data coming from  
different types of observation. In the following, a summary of 
the data available for NGC~3132 is presented. 
 
The logarithm of the observed H$\beta$ flux ranges from -10.49
(\cite{P71}) to -10.20 (\cite{P77}). The extinction E(B-V) is of the order 
of 0.1 (\cite{P77}, Mendez 1978, Feibelman 1982, Gathier, Pottasch \& 
Pel 1986).
 
The calculated values for the Zanstra \ion{He}{2} temperature of the
ionizing star 
of NGC~3132 ranges from 73 000 K (de Freitas Pacheco, Codina \& Viadana 1986) 
to 110 000 K (\cite{P96}).
Distance determinations indicate values from
$\sim$ 0.51 kpc (\cite{DF86}, Pottasch 1996) to 1.63 kpc 
(Torres-Peimbert \& Peimbert 1977). The ionizing star 
luminosity available in the literature is 72 L$_{\odot}$ (Mendez 1978) or 
$\sim$ 125 L$_{\odot}$ (\cite{P84}).

HST/WFPC2 images for \halpha/, \forb{O}{1}{6300}, \forb{O}{3}{5007}, 
\forb{N}{2}{6583}, and \dforb{S}{2}{6717}{6731} are available at the CADC  
server from the Trauger's  proposal \# 6221\footnote{Guest User, Canadian 
Astronomy Data Center, which is operated by the  
National Research Council, Herzberg Institute of Astrophysics, Dominion 
Astrophysical Observatory}. Lower resolution images and contours were 
obtained by \cite{J88}, \cite{B90}, \cite{P90}, 
\cite{C81}, and \cite{S92}.
  
Spectroscopic observations of NGC~3132 can be found in \cite{A64}, 
\cite{K76}, \cite{T77}, and \cite{B90}. 
Expansion velocities of 14.7 \kms/ for \forb{O}{3}{5007} and 21.0
\kms/ for 
\dforb{O}{2}{3726}{3729} were obtained by Meatheringham, Wood \& Faulkner 
(1988), for slits along 
the major axis. High resolution spectra for \forb{O}{3}{5007} in five 
positions were obtained by \cite{SD86}, who also deduced 
a velocity of 14 \kms/ at the central position. 
The line profiles are all asymmetrical and double peaked.  
Since the observed red peak is more intense towards the center,
the asymmetry can not be due to local absorption.

Electronic densities are generally obtained from the [\ion{O}{2}] or
[\ion{S}{2}] line  
intensity ratios corresponding to intensities integrated on a slit. 
\cite{T77} obtained 1000 cm$^{-3}$, while \cite{M88} and 
Stanghellini \& Kaler (1989) give, respectively, 600 cm$^{-3}$ and 430 
cm$^{-3}$.
More detailed observations show a variation 
of the electronic density measured in different positions.
Density determination by \cite{J88}, from the \dforb{S}{2}{6717}{6731} 
line intensity ratio, shows
a double-peaked distribution along the NS direction, 
reaching  up to 1300 cm$^{-3}$ at the periphery of the nebula and 
decreasing to 300 cm$^{-3}$ at the central position.  
Recent observations have confirmed this density distribution trend 
(Monteiro, Gruenwald \& de Souza, in preparation). 
{\it It will be shown that the decrease in density at the central 
region is the key observation that rules out the description 
of NGC~3132 as a geometrical ellipsoidal shell.} 

\section{The ellipsoidal model} 
\label{sec-bass} 

The 3D photoionization code used in this paper 
is described in Gruenwald, Viegas \& Brogui\`ere (1997). 
The output of the code is  
treated by IDL tools performing rotations, projections, and  
velocity profiles. A  full description of these tools is given by  
MGV2000.  

In order to fit the observed morphology of NGC~3132 (\forb{O}{3}{5007} 
and \forb{N}{2}{6583} line images), as well as the emission line spectra taken 
through 4 slits, \cite{B90} proposed a model obtained
by a 3D photoionization code.
The geometry and input parameters proposed 
by \cite{B90} are taken as a starting point for our 3D model 
of NGC~3132. Elements heavier than Ne were not included in \cite{B90} model.
For S, Cl, and Ar we assumed a solar abundance (\cite{GA89}), while Mg, S, Cl, 
and Fe are depleted by one hundred (Stasinska \& Tylenda 1986).
Thus, the gas chemical abundance, in number, is:  
H: 1.0, He: 0.126, C: 7.1(-4), 
N: 2(-4), O: 6(-4), Ne: 8.2(-5),  
Mg: 3.8(-7), Si: 3.55(-7), S: 1.62(-5), Cl: 3.16(-7), Ar: 3.6(-6), Fe: 
4.7(-7).  
The central ionizing radiation is a black-body of 90~000~K 
and 150~L$_\odot$. 
The model consists of two prolate ellipsoids defining a 
shell and an inner cavity (the outer zone being empty).  
The inner ellipsoid has a semi-minor axis of 1.27 $10^{17}$cm and a 
semi-major axis of 2.32 $10^{17}$cm, while the outer one has  
1.96 $10^{17}$cm and 2.45 $10^{17}$cm, respectively. 
The  density in the inner cavity is constant (464 
cm$^{-3}$). In \cite{B90}, the shell density varies along the three directions.
Here only the most important density gradient (along the major axis Z) 
is assumed. It varies from $\sim$ 1200
cm$^{-3}$ at one pole to $\sim$ 900 cm$^{-3}$ at the opposite 
pole. This gradient is needed in order to reproduce
the observed asymmetrical brightness enhancement 
shown by the optical images. 

The calculations are performed assuming the on-the-spot  
approximation (see  MGV2000).
Initially the cube containing 1/8 the nebula is divided 
in $36^3$ cells. In order to 
improve the numerical resolution, some 
of these cells are subdivided, resulting in 133 000 cells.   
Once the 
calculations converged, the whole nebula is recovered (see MGV2000).
Following \cite{B90}, the images are obtained after a 7\degre/.5
rotation around the X axis  
(perpendicular to the line of sight). 

\subsection{Matter- and Radiation-bound models}
\label{sub-bas-MRmod} 
The results obtained by \cite{B90} for the line intensities are not 
reproduced by our ellipsoidal model. 
First, the geometrical thickness of the shell, adopted by these authors, 
is not large enough to absorb all the photons. 
Our model with the \cite{B90} geometry leads to a matter-bound (M-bound) 
nebula. The model is then unable 
to reproduce the observed \dforb{N}{1}{5198}{5200} strong lines,
characteristic of a radiation bound nebula.
Notice that the measured intensity for \forb{O}{1}{6300}, $\sim$ 0.3 \hbeta/  
(Torres-Peimbert \& Peimbert 1977), also indicates that the nebula must
be mainly radiation-bound (R-bound). 
Although the details of the \cite{B90} 
code are not available in their paper, the difference with our results 
may come from the optical depth calculation, which is the 
key-parameter determining the model. Models with a low number 
of cells (which implies too large cells) tend to overestimate the optical 
depth. In this case, for the same  
geometrical thickness, the model would be R-bound instead of M-bound.
Our model has 40 times more cells than \cite{B90}.
This can explain why, using the same input parameters, our model is M-bound 
while theirs is R-bound.
In order to have a R-bound shell,  
reproducing the observed low-ionization emission lines,   
the outer ellipsoid size must be increased by at least 20\%.
A second reason for the discrepancies between our model and that of
\cite{B90} is the lack of elements heavier than Ne in their 
code, which leads to 
an underestimation of the cooling processes. In order to test this 
effect, a heavy-element free-\cite{B90} model was  
built. The \forb{O}{3}{5007} and \forb{N}{2}{6583} line intensities
(relative to \hbeta/) increase, respectively, by 30$\%$ and 16$\%$.
This can partly explain the differences with \cite{B90} results, 
since the optical depth effect, discussed above, must be also important.  

In the following, we will refer to the
matter-bound model using the dimensions of \cite{B90} as the 
M-\cite{B90} model and to the 20$\%$ ``extended'' radiation-bound model as the 
R-\cite{B90} model. The results of both models  
are presented in Table \ref{tab-obs}. As expected, the 
stronger differences 
appear for the low ionization lines (\dforb{N}{1}{5198}{5200} and 
\forb{O}{1}{6300}). Nevertheless, 
even  the R-\cite{B90} model is far from reproducing the results 
obtained by \cite{B90}, as discussed below.  
\placetable{tab-obs}

\subsection{Ellipsoidal model results}
\label{sub-bas-res}
The superposition of the \forb{N}{2}{6583} HST image and the 
theoretical isophotes obtained from the R-\cite{B90} model 
is shown in Fig. \ref{fig-hst-b90}. 
The shape is well reproduced, as well as the variation of 
the surface brightness along the major axis. To adjust the modeled 
shape and dimensions to the observed ones, a reduction of 4$\%$  
in the distance used by \cite{B90} (670 pc) is necessary. Note that 
this new value for the distance is in the range of the calculated distances
to NGC~3132 (\S 2) 

For a comparison of our results with the observations of \cite{J88}, 
the theoretical N-S  
density distribution is calculated from the theoretical 
\dforb{S}{2}{6717}{6731} ratio (using the \cite{AB97} fit)
for a 4" width slit, centered on the nebula.
As seen from Fig. \ref{fig-sii-dens}, this model assuming an 
ellipsoidal geometry (dashed line) does not reproduce the observed central  
decrease in density.

In order to obtain the theoretical \forb{O}{3}{5007} emission line 
profiles,  
a radial velocity law 
$\vec V  = \alpha . \vec r / r + \beta . \vec r $  
is used.
Three possible sets of $\alpha$ and $\beta$ values are assumed:
$\alpha = 14.7, 9.1, 0$ \kms/ and 
$\beta = 0, 9.1/r_o, 23.8/r_o$ km s$^{-1}$ cm$^{-1}$, respectively. 
The normalization factor, $r_o$ = 3 10$^{17}$cm, is of 
the order of the size of the nebula.
The \mvelo{5007} and \mvelo{3726} emission line mean velocities 
for the whole nebula (as defined in MGV2000) are given in Table
\ref{tab-velo}.
For the three sets of values defining the velocity law, the parameters 
were chosen in order to give  
\mvelo{5007} = 14.7 \kms/, as obtained 
from the observed profile reported by \cite{M88}.  The 
corresponding theoretical \mvelo{3726} mean velocity  
depends on the parameters $\alpha$ and $\beta$ and is always 
slightly lower than the observed value (21 \kms/).  
The five positions at which the \forb{O}{3}{5007}
line profiles were observed are shown in \cite{SD86} (their Fig. 1a)
superposed
on an U-band image (Kohoutek \& Lausten 1977).
The calculated velocity profiles for \forb{O}{3}{5007} 
are shown in Fig. \ref{fig-velo-b90}. 
Since the three sets of parameters of the adopted velocity law 
give similar line profiles, only the results for the $\alpha = 9.1$ \kms/ 
and $\beta = 9.1/r_o$ km s$^{-1}$ cm$^{-1}$ law is shown.
In the upper-right panel the aperture is shifted from the center
by 2.8'' NE. This 
shift corresponds to the pointing accuracy of the 
telescope used by \cite{SD86} and it allows to check if 
the effect of the pointing precision on the profile is important.
The central double peak is reproduced. However,
the observed asymmetry can only be obtained if the nebula is 
tilted by about 30\degre/, whereas the results shown in 
Fig. \ref{fig-velo-b90} were obtained assuming a rotation angle of
7\degre/.5, as proposed by BDG90. 
The observed double peaks at the four external  
positions are not reproduced. Note that the observed profiles were taken by SD86 
at positions where the U band emission is strongest. In the U band, the 
nebular emission is dominated by the \forb{O}{2}{3726} line, so the 
observations were taken 
at the extreme outer part of the nebula. 
Thus, the radial velocity at these 
positions is quasi-perpendicular to the line of sight and an 
apparent velocity of $\sim$ 10 \kms/ would require an unrealistic 
high radial velocity. 

\placetable{tab-velo}


Finally, a simple way to verify the inadequacy of the ellipsoidal model 
in reproducing the density gradient and the low ionization lines is the 
following: let us consider a dense shell ($\sim$ 1000 cm$^{-3}$) with a
geometrical thickness $t$ and 
radius $R$, around a less dense cavity ($\sim$ 300 cm$^{-3}$). 
The line emissivity is proportional to the square of the density. Thus,   
in order to reproduce 
the observed density distribution, the main contribution to the 
central emission must come from the low density gas, implying $t < 0.02 R$.  
Such a thin shell will have a total emissivity far below the observed 
one, and will be matter-bound, not producing the  
low ionization lines. A radiation-bound model, necessary to explain the low 
ionization lines cannot reproduce the density gradient as shown above. 
This rules out the description of NGC~3132 as a 
{\it closed ellipsoidal shell}.  
An ellipsoidal shell with a hole just aligned to the 
central line of sight could be proposed, but such an {\sl ad hoc} 
geometry seems unrealistic.  

From all the results presented above we conclude that 
an ellipsoidal model for NGC~3132 is not appropriate to explain the 
observational data. Thus, a new model is proposed below.
 
\section{The \diab/ model} 
 \label{sec-diab}

PNe imaging suggests that several objects can have a ``Diabolo'' shape (e.g.
Hour-Glass nebula and NGC 2346). 
In order to reproduce this shape, we adopt a two-zone 
gas distribution, illustrated in Fig. \ref{fig-densh4}: a) a dense
zone (with a density of 1300 cm$^{-3}$)  
defined by the intersection of two spheres of the same radius  
(3. 10$^{17}$ cm) but with centers shifted by 10$^{17}$ cm, defining 
the ``Diabolo-shape'', and b) a low-density gas, 300 cm$^{-3}$, 
filling the rest of the nebula. The shadowing in the figure is just used  
to give a better 3D visualization and does not correspond to a density 
variation. 
The model was calculated for 1/8 of
the nebula and the whole nebula was reconstructed in a 10$^6$ cells
cube.
We keep the chemical abundances and ionizing star characteristics of 
\cite{B90} model (\S \ref{sec-bass}). No attempt to adjust these 
parameters were made in order to improve the fit of the emission line
ratios. In fact, the scope of this paper is just to show that a strong 
constraint on the geometry is the density distribution. 
Once this distribution is reproduced,  other observed features, 
unexplained by previous models, can also be explained.
 
\subsection{The Diabolo model results}
\label{sub-diab-res}

The effect of varying the angle of view of the nebula is shown in 
Fig. \ref{fig-diab-rot}.
Practically all the observed regular morphologies of PNe are reproduced:
from a clearly butterfly image (axis of symmetry
perpendicular to the line of sight) to a well round
nebula (axis of symmetry aligned to the line of sight) through an ellipsoidal
one.
With an appropriate orientation of 40\degre/ relative to the line
of sight, the observed optical shape of NGC~3132 (ellipsoidal shape) can be
reproduced. As said above (\S 1), such feature can also be reproduced by 
the ellipsoidal shell model of Zhang \& Kwok (1998), as well as by a model
similar to that of Masson (1990). However, these models are not 
self-consistent with respect to the ionization balance and, as will be shown 
below, other constraints will rule out an ellipsoidal model for this nebula.


With a \diab/ morphology the contribution from the dense region  
to the central emission is reduced, helping to solve the close ellipsoidal 
shell problems.
At the same time, the projected velocity of the gas in the
external parts of the nebula will be high enough, reproducing 
the observed double peak, as shown below.  

Regarding the emission line ratios, the results of the \diab/ model 
(fourth column of Table \ref{tab-obs}) are similar to 
those obtained from the R-\cite{B90} model. In addition, 
HST line images in H$\beta$, \forb{O}{3}{5007}, 
\forb{N}{2}{6583} and \forb{O}{1}{6300} are shown 
Fig. \ref{fig-ima-diab} (four upper panels) and can be compared to
the corresponding \diab/ model images (four lower panels).
The observed  ellipsoidal shape is well  
reproduced, as well as  the ionization  
stratification. The brightness asymmetry along the major axis
is not reproduced by the adopted axi-symmetrical 
model of the nebula. Such asymmetry could be explained either by a 
density gradient
or an ionization source that is not in the geometrical center of the Diabolo.


The electronic density distribution obtained from the
\dforb{S}{2}{6717}{6731} ratio is also shown in
Fig. \ref{fig-sii-dens} (solid line).  
{\it The observed decrease of the density (\S 2) in the central part 
of the nebula is reproduced}.
This and the line profiles discussed below are the main improvements 
obtained by the Diabolo shape.

The same velocity laws applied for the R-BDG90 model are used (\S3.2) 
with $r_o$ = 5 10$^{17}$cm.
The three different \forb{O}{3}{5007} profiles obtained with the  
three sets of parameters of the velocity 
law are shown in Fig. \ref{fig-prof-diab} and can be compared to 
the observations (\cite{SD86}).  
The profiles show the double peaks   
at the six positions, {\it a feature not reproduced by the ellipsoidal shell 
model}. 
The asymmetry is well reproduced at the four external positions.
The asymmetry at the central position is not reproduced, as long as 
an axi-symmetric nebula is used, even considering a shift of 2.8'' 
as shown in the upper right panel. 
Notice that at the central position only the low
density gas contributes to the emission. Thus, any asymmetry in 
the velocity field and/or in the low
density distribution will lead to the observed asymmetry, requiring no 
change in the higher density distribution associated to the \diab/ shape.

The three sets of parameters for the velocity law were chosen in order to 
obtain the same \mvelo{5007} (14.7 \kms/, Table \ref{tab-velo}; the results 
for \mvelo{3726} remain the same as for
the R-BDG90 models); however, for $\alpha$ = 0 (dashed  lines), 
the peaks of the observed 
profiles at the central part of the nebula occur for a lower 
velocity (Fig. \ref{fig-prof-diab}, upper panels).
Furthermore, in this case, the velocity law 
does not reproduce the observed profiles since one of the peaks is almost 
inexistent.
The results obtained with the other two sets of parameters (solid and 
dot-dashed lines) are very similar. Nevertheless, for $\alpha$ $\not=$ 0 and 
$\beta$ $\not=$ 0, \mvelo{3726} is higher than \mvelo{5007}, approaching the 
observed result (Table \ref{tab-velo}) of \cite{M88}.

\section{Concluding remarks} 
\label{sec-concl}

We have shown in this paper that the lack of high resolution observations 
(spatial and spectroscopic) may lead to a wrong conclusion about 
the morphology of a PN.

A new modeling tool based on a 3D photoionization code is used to 
consistently reproduce all the available observations of NGC~3132. 
Using an ellipsoidal geometry, all the low resolution observations  
(flux, emission line spectrum, imaging) could be reproduced.
However, neither the density variations indicated by
the \dforb{S}{2}{6717}{6731} ratio, nor the 
\forb{O}{3}{5007} velocity profiles could be explained by such a model.  
To account for these observations, a drastic change in the geometry of 
the nebula is necessary. The proposed Diabolo shape offers the 
solution to the density distribution, as well as to the asymmetric 
double-peaked emission line profiles, while also explaining the low
resolution observations.

A Diabolo shape may reproduce the observed bipolar morphology of many PNe. 
In fact, Bryce, Balick \& Meaburn (1994) suggested that NGC~6720 
is not a spherical expanding shell as often assumed, but must be bipolar in 
nature, since the observed [NII] emission line profiles are split in two 
components.
A bipolar shape emerges from studies of the formation and 
evolution of PNe, in particular those
related to the interacting wind model (Kwok, Purton, \& Fitzgerald 1978). 
Recent hydrodynamical models (e.g. Dwarkadas, Chevalier \& Blondin 1996;
\cite{M97}) 
suggest that a Diabolo type shape 
can be produced in the accepted scenario of PNe formation
(post AGB envelope ejection with wind interaction). In fact,
recent high resolution images taken with the Hubble Space Telescope show 
that PNe with this type of structure is not uncommon. Some nice examples 
are the Hour-Glass nebula, NGC~2346, NGC~6537 and Hb-5.

No attempt was made to fine tuning the emission line ratios. This 
can be achieved through changes in the ionizing star characteristics, in 
the chemical abundances, as well as through small adjustments in the density 
distribution, which will not change the main conclusions of this paper, 
in particular the Diabolo shape.
Furthermore, in the models presented here, the  
ionizing source is assumed to be at the center of symmetry of the nebula. 
However, there are evidences that the star should be shifted 
by $\sim$ 1.7'' from the geometrical center of the nebula 
(Kohoutek \& Lausten 1977 and \cite{SD86}).
A model accounting for this 
out-of-center ionizing star may still improve the results, leading 
to a better agreement with the observations regarding the  
brightness asymmetry and the emission line profile at the central 
position. 

We have shown here the importance of a 3D photoionization model, combined 
with  imaging results and also with high resolution observations, to impose
relevant constraints on the PN geometry. 
The misclassification of NGC~3132 as an ellipsoidal PN instead
of a bipolar one casts doubts on the 
statistical results correlating the morphology and other PNe
properties, such as abundances or progenitor type.

\acknowledgements
The authors would like to thank the referee, Dr. Sun Kwok, for his 
valuable comments and suggestions. 
C. Morisset is thankful to IAGUSP for the hospitality during his 
post-doctoral visit.
This work was supported by CNPq (n.~304077/77-1, n.~306122/88-0, and 
n.~150162/96-0), FAPESP (n.~97/13428-4 and n.~98/01922-7), and 
PRONEX/FINEP (n.~41.96.0908.00).

\newpage
\figcaption[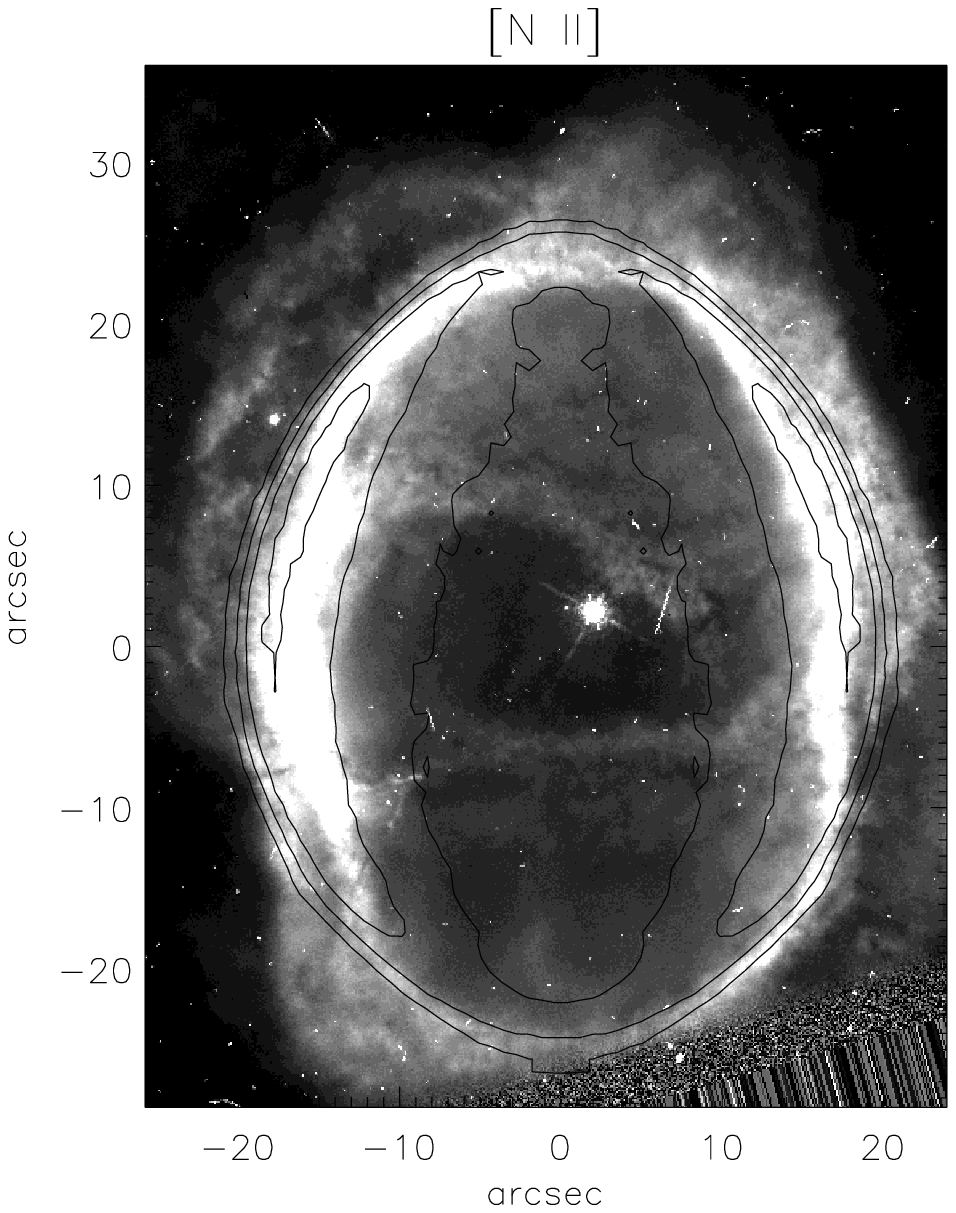]{Superposition of the HST and the theoretical images
of \forb{N}{2}{6583}. The theoretical results correspond to the
R-\cite{B90} model.
\label{fig-hst-b90}}  

\figcaption[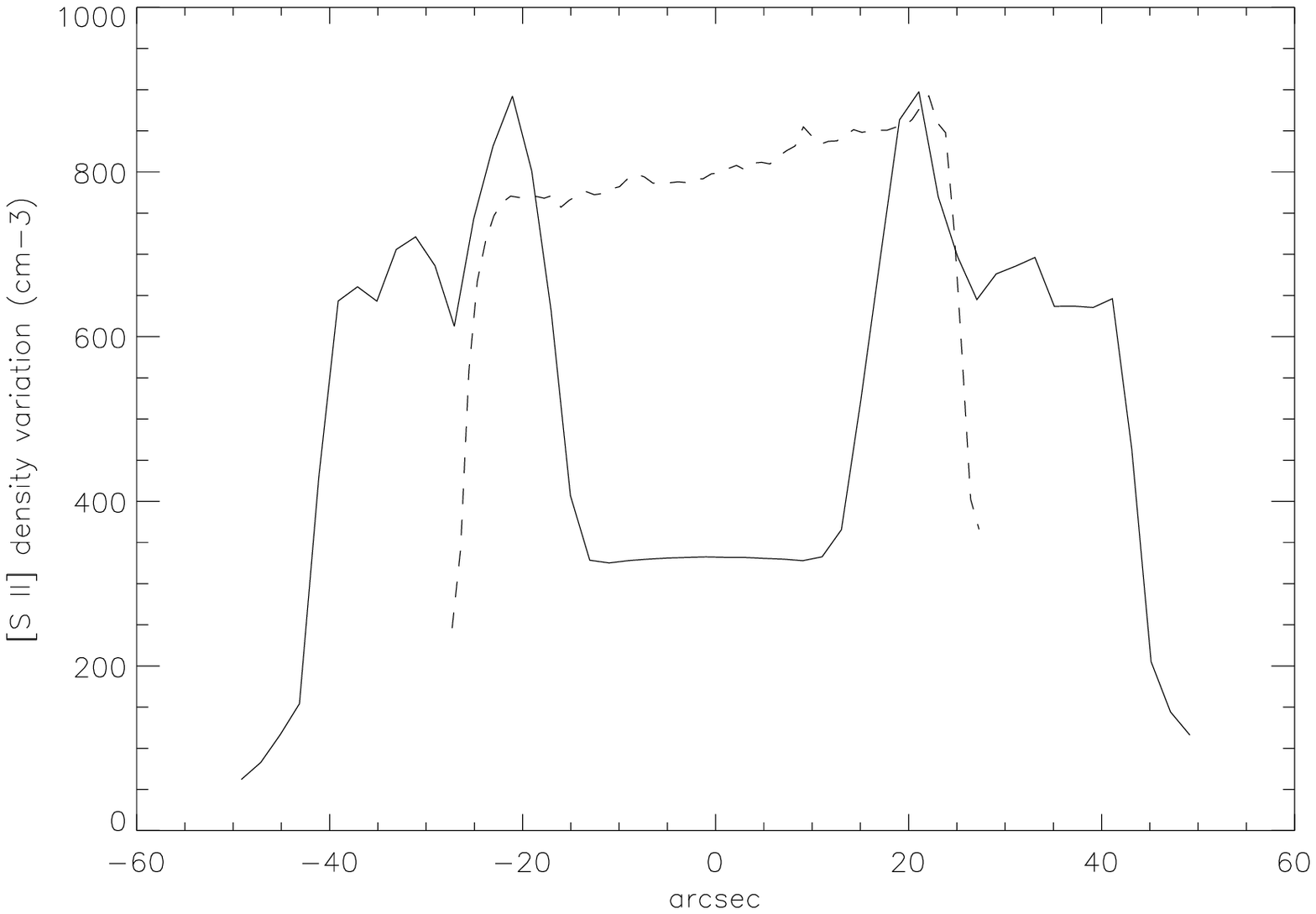]{N$\leftrightarrow $S variation of
\dforb{S}{2}{6717}{6731} density from the R-\cite{B90} model (dashed
line) and from the \diab/ model (solid line). 
\label{fig-sii-dens}} 

\figcaption[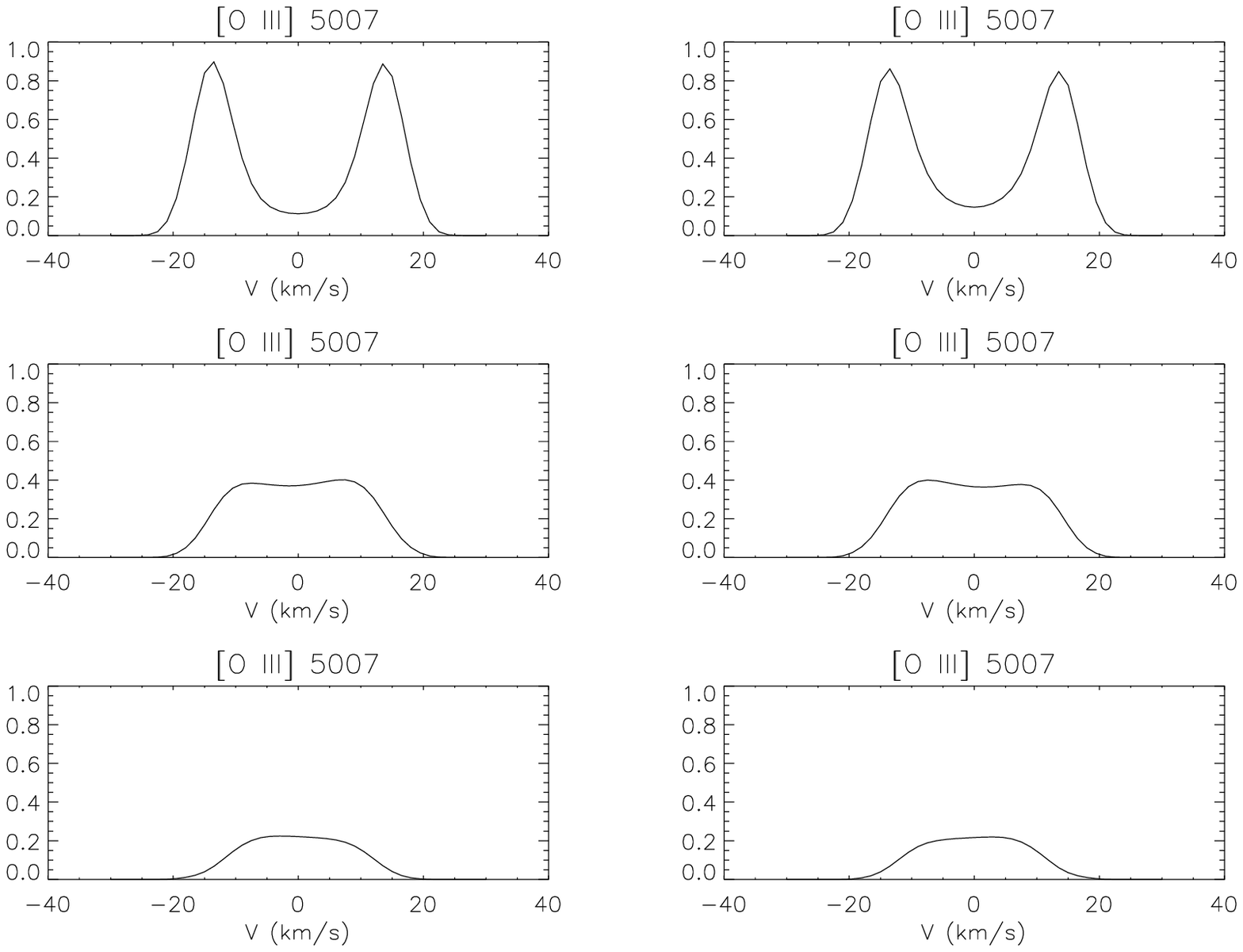]{\forb{O}{3}{5007} velocity profiles obtained from the
R-\cite{B90} model. The positions and the size aperture are the same as
in \cite{SD86} (their Fig. 2). In the upper-right panel the aperture 
is shifted from the center by  2.8'' (see text).
\label{fig-velo-b90}}  

\figcaption[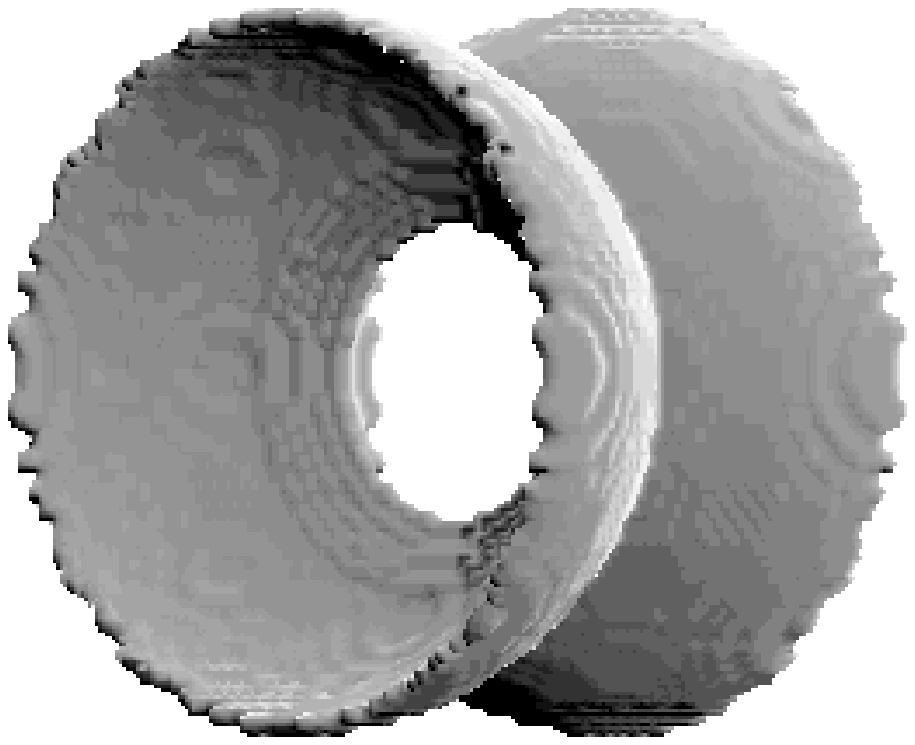]{Gas distribution for the \diab/ model. Only the
denser zone (1300 cm$^{-3}$) is shown. 
\label{fig-densh4}}  

\figcaption[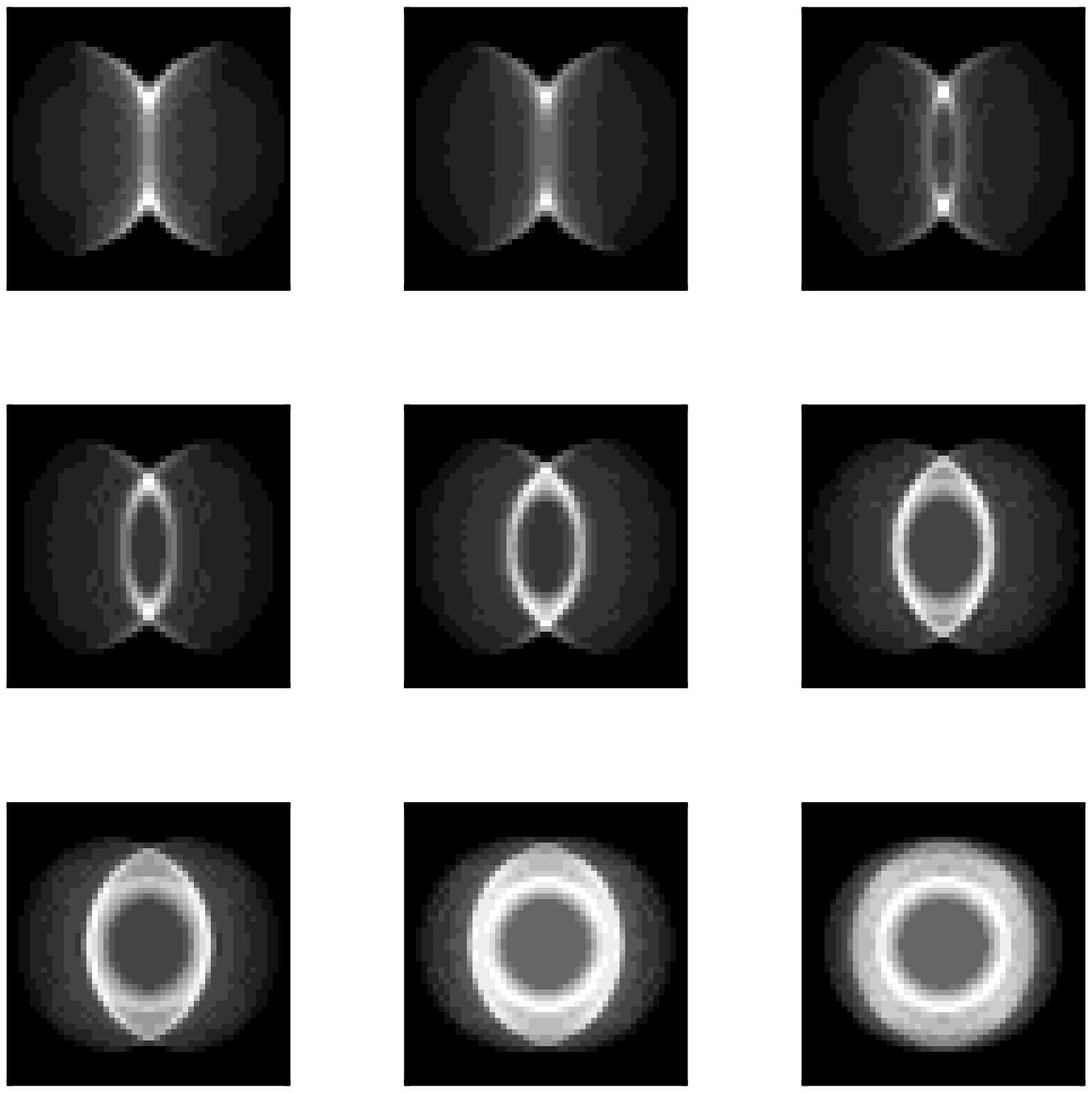]{\hbeta/ images from the \diab/ model. 
From the upper-left to the lower-right panel the 
angle between the axis of symmetry and the sky plane increases by 10\degre/
from 0\degre/ to 80\degre/.
\label{fig-diab-rot}}

\figcaption[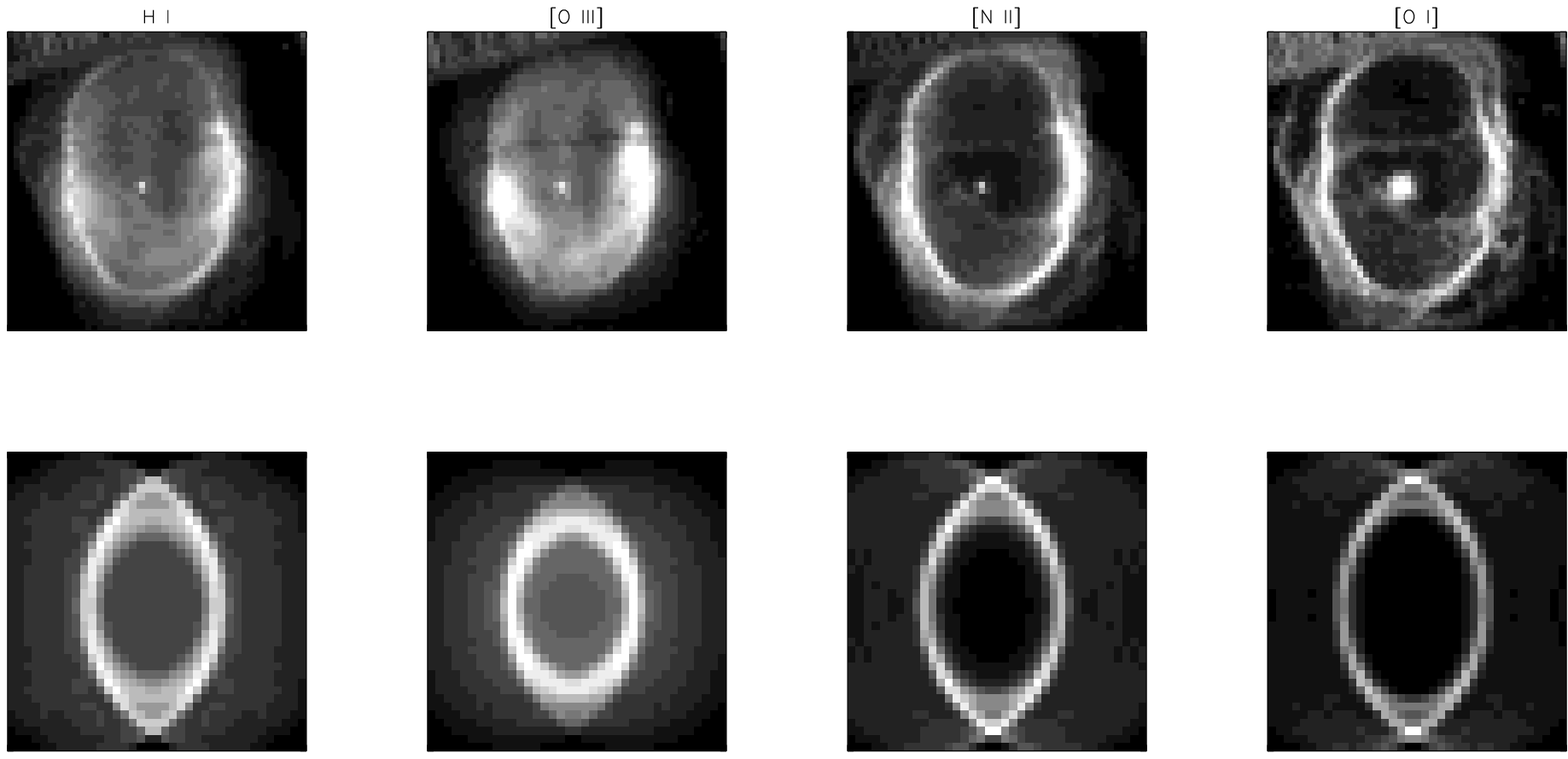]{ The top panels correspond to
HST images of NGC~3132 while the botton panels to the 
\diab/ model. 
\label{fig-ima-diab}}  

\figcaption[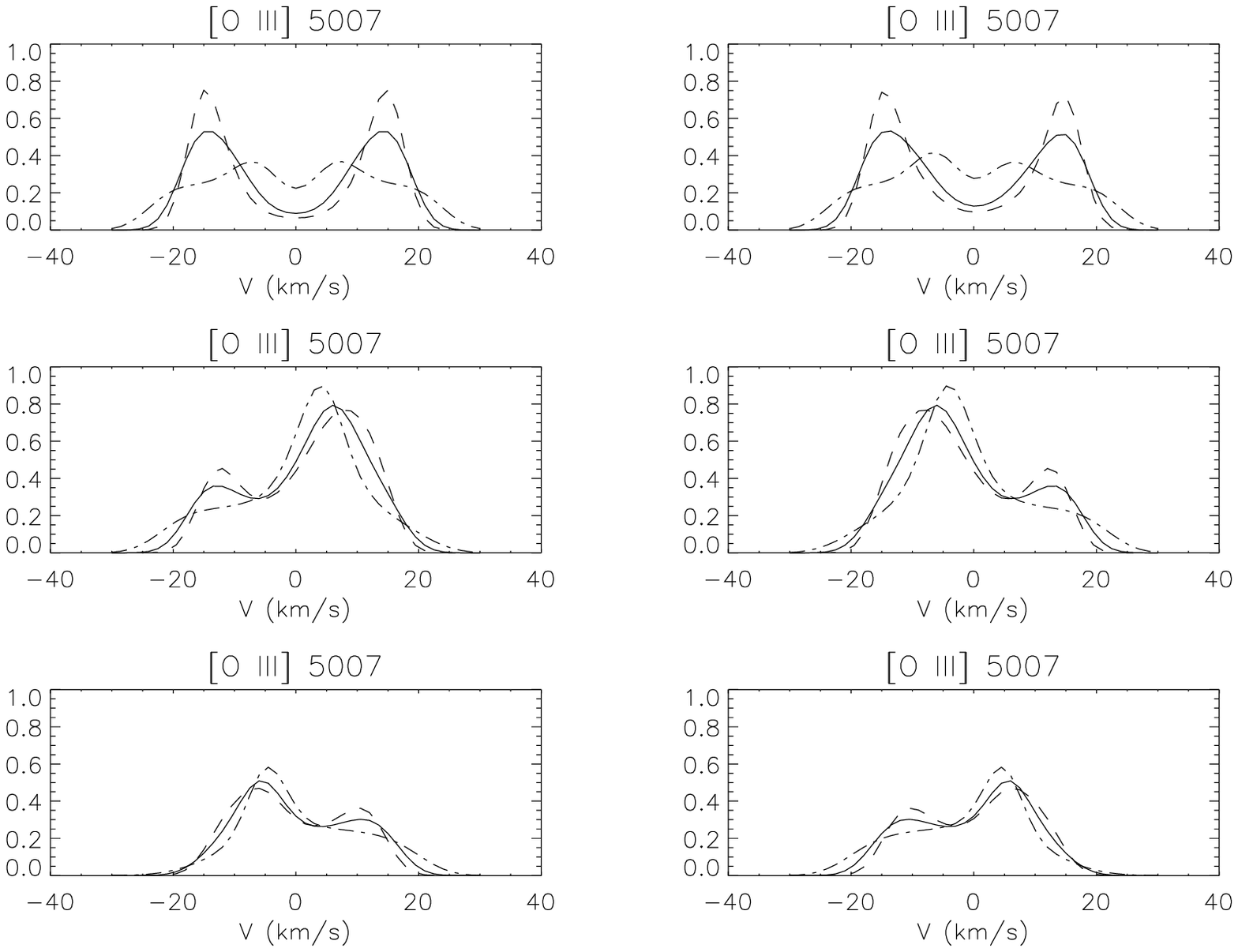]{\forb{O}{3}{5007} velocity profiles for the
\diab/ model using the velocity law 
$\vec V  = \alpha . \vec r / r + \beta . \vec r $, with r$_0 = 3 10^{17}$cm, 
and
($\alpha$, $\beta$ ) = (14.7, 0), (9.1, 9.1/r$_o$), and (0, 23.8/r$_o$) are 
shown, respectively, by dashed, solid, and dot-dashed lines.
\label{fig-prof-diab}} 

\newpage
 
\begin{table}[h] 
\caption{Emission line intensities relative to  \hbeta/  \label{tab-obs}} 
\begin{tabular}{rccc} 
\tableline 
Line \AA & M-\cite{B90} & R-\cite{B90} & \diab/\nl   
\tableline 
\forb{O}{3}{4363} 	& 0.022	& 0.018	& 0.013	\nl  
\allo{He}{2}{4686} 	& 0.085	& 0.059	& 0.065	\nl  
\forb{O}{3}{5007} 	& 6.045	& 4.700	& 3.690	\nl  
\forb{N}{1}{5200} 	& 0.010	& 0.159	& 0.200	\nl  
\allo{He}{1}{5876} 	& 0.175	& 0.189	& 0.190	\nl  
\forb{O}{1}{6300} 	& 0.037	& 0.320	& 0.353	\nl  
\forb{N}{2}{6583} 	& 2.792	& 5.090	& 5.337 \nl  
\forb{S}{2}{6717} 	& 0.193	& 0.689	& 0.825	\nl  
\forb{S}{2}{6731} 	& 0.236	& 0.745 & 0.846	\nl  
H$\alpha $        	& 2.904	& 2.915	& 2.917	\nl  
H$\beta^1 $       	& 6.553	& 9.651	& 9.050	\nl 
\tableline 
\end{tabular} 
 
$^1$ {\footnotesize In $10^{-11}$ \flux/, for a distance of 670 pc.}  
\end{table} 

\begin{table}[h] 
\caption{Mean line velocities for the adopted velocity law \label{tab-velo}} 
\begin{tabular}{rrcc} 
\tableline 
$\alpha^a$& $\beta^b$  & \mvelo{5007}$^a$ & \mvelo{3726}$^a$\nl  
\tableline 
14.7 & 0. & 14.7 & 14.7 \nl 
9.1 & 9.1 & 14.7  & 15.9 \nl 
0. & 23.8 & 14.7 & 17.8 \nl 
\tableline 
\end{tabular} 
 
$^a$ {\footnotesize in \kms/} 
 
$^b$ {\footnotesize in \kms//r$_0$, with r$_0 = 3 10^{17}$cm for the R-BDG90 
model and  r$_0 = 5 10^{17}$cm for the Diabolo model} 

\end{table} 
\newpage
\setcounter{figure}{0}
\begin{figure} 
\plotone{fig1.ps}
\caption{}
\end{figure} 

\begin{figure} 
\plotone{fig2.ps} 
\caption{}
\end{figure} 

\begin{figure} 
\plotone{fig3.ps} 
\caption{}
\end{figure} 

\begin{figure} 
\plotone{fig4.ps} 
\caption{}
\end{figure} 

\begin{figure} 
\plotone{fig5.ps} 
\caption{}
\end{figure} 
  
\begin{figure} 
\plotone{fig6.ps}
\caption{}
\end{figure} 
 
\begin{figure} 
\plotone{fig7.ps} 
\caption{}
\end{figure}

\end{document}